\documentclass[a4paper,11pt]{article}
\usepackage{jheppub} % for details on the use of the package, please
\usepackage{amsmath,epsfig}
\usepackage{amssymb,amsfonts}
\usepackage{latexsym}
\usepackage[latin1]{inputenc}
\usepackage[T1]{fontenc}
\usepackage[english]{babel}
\usepackage{slashed}
\usepackage{empheq}
\numberwithin{equation}{section}
\usepackage{dsfont}
\usepackage{cancel}
\usepackage{overpic}
\usepackage{subcaption}
\usepackage{accents}
\usepackage{subeqnarray}
\usepackage{xcolor}

\usepackage{graphicx}
\usepackage{physics}
\usepackage{caption}
\usepackage{subcaption}
\usepackage{comment}
\usepackage{mdframed}
\usepackage{lipsum}
\usepackage{lmodern}  

\usepackage{longtable}
\usepackage{color}
\usepackage{tensor}
\usepackage[mathscr]{euscript}
\usepackage{physics}
\usepackage{overpic}
\usepackage{multicol}

\newcommand{\exclude}[1]{}
\def\d{\mathrm{d}}

\def\a#1{\alpha_{#1}}

\def\bea{\begin{eqnarray}}
	\def\eea{\end{eqnarray}}

\def\Re{\textrm{Re}}

\def\2b2[#1,#2][#3,#4]{\left( \begin{array}{cc} #1 & #2 \\ #3 & #4 \end{array}
	\right)}
\def\3b3[#1,#2,#3][#4,#5,#6][#7,#8,#9]{\left( \begin{array}{ccc} #1 & #2 #3 \\
		#4 & #5 & #6\\#7&#8&#9\end{array} \right)}

\newcommand\fverb{\setbox\pippobox=\hbox\bgroup\verb}
\newcommand\fverbdo{\egroup\medskip\noindent%
	\fbox{\unhbox\pippobox}\ }
\newcommand\fverbit{\egroup\item[\fbox{\unhbox\pippobox}]}

\newcommand{\bear}{\begin{eqnarray}}
	
	\newcommand{\eear}{\end{eqnarray}}

\newcommand{\bsea}{\begin{subeqnarray}}
	\newcommand{\esea}{\end{subeqnarray}}
\newbox\pippobox

\def\d{\delta}

\def\6{\partial}
\def\a{\alpha}

\def\m{\mu}
\def\n{\nu}

\def\sq
\def\a{\alpha}
\def\b{\beta}

\def\d{\delta}

\title{\textbf{Hydrostatic equilibrium in multi-Weyl semimetals}}

\author[a,b]{Jewel Kumar Ghosh,}
\author[c]{Francisco Pe\~na-Ben\'itez,}
\author[d,e]{and Patricio Salgado-Rebolledo}

\affiliation[a]{Department of Physical Sciences, Independent University, Bangladesh (IUB), Bashundhara RA, Dhaka 1229, Bangladesh}

\affiliation[b]{Center for Computational and Data Sciences (CCDS), Independent University, Bangladesh, Dhaka 1229, Bangladesh}

\affiliation[c]{Institute of Theoretical Physics, Faculty of Fundamental Problems of Technology, Wroc\l{}aw University of Science and Technology, 50-370 Wroc\l{}aw, Poland}

\affiliation[d]{Asia Pacific Center for Theoretical Physics (APCTP), Pohang, Gyeongbuk 37673, Korea}

\affiliation[e]{Instituto de Ciencias Exactas y Naturales (ICEN), Universidad Arturo Prat,\\
Playa Brava 3256, 1111346 Iquique, Chile
\vspace{.3cm}}

\emailAdd{jewel.ghosh@iub.edu.bd} %\emailAdd{francisco.pena-benitez@pwr.edu.pl}
\emailAdd{salgado.rebolledo@apctp.org}

\abstract{We study the hydrostatic equilibrium of multi-Weyl semimetals, a class of systems with Weyl-like quasi-particles but anisotropic dispersion relation $\omega^2 \sim k_\parallel^2 + k_\perp^{2n}$, with $n$ a possitive integer. A characteristic feature of multi-Weyl systems is the lack of Lorentz invariance, instead, they possess the reduced spacetime symmetry  $(SO(1,1)\times SO(2))\ltimes \mathbb R^4$. In this work we propose a covariant formulation for the low energy theory, allowing for a minimal coupling of the fermion field to external geometric background and $U(1)$ gauge field. The non-Lorentzian structure of the field theory demands introducing an Aristotelian spacetime analogous to the so-called stringy Newton-Cartan geometry \cite{Andringa:2012uz}.  Our proposal allows for a systematic study of the hydrostatic properties via the derivation of the partition function of the system. In addition to multi-Weyl models, our formulation can be applied to systems with similar spacetime symmetry groups, such as Bjorken flow.
}

\keywords{Weyl fermions, multi-Weyl semimetals, non-Lorentzian geometry, non-boost invariant hydrodynamics, Hydrostatic partition function. \\ \ \\ \ \\ \ \\  \textit{Dedicated to the memory of Umut G\"ursoy}}

\begin{document} 
\maketitle
\flushbottom

\section{Introduction}\label{sec:Intro}

Multi-Weyl semimetals (mWSMs) have emerged as a fascinating class of topological materials characterized by anisotropic energy dispersion and higher-order topological charge at their Weyl nodes and are expected to be realized in realistic materials such as $\mathrm{HgCr}_2\mathrm{Se}_4$ and $\mathrm{SrSi}_2$ \cite{Xu2011ChernHgCr2Se4,Fang2012Multi-WeylSymmetry,Huang2016NewFermions}. Unlike conventional Weyl semimetals \cite{Gorbar:2021ebc}, where the quasiparticles exhibit linear dispersion in all momentum directions, mWSMs display quadratic or cubic dispersion in certain directions, leading to unique transport and optical properties under external perturbations, such as external electromagnetic fields \cite{Dantas2018MagnetotransportApproach,Dantas2020Non-AbelianSemimetals,SubramanyanGeometricSemimetals,Ghosh2023ElectricStrain,Raj2023PhotogalvanicSemimetals, Medel:2024wcj}, and strain \cite{SubramanyanGeometricSemimetals,Juricic2024HolographicDislocations,SukhachovElectronicSystems}.
Furthermore, understanding the interaction between topology and geometry in these materials requires a careful treatment. In fact, pure gauge anomalies for multi-Weyl fermions are relatively well understood \cite{LeporiLucaandBurrello2018AxialSemimetals}. However, the role of gravitational anomalies is a problem that to our knowledge, has not been studied \footnote{See \cite{Dantas2020Non-AbelianSemimetals} for a discussion on the possible gravitational anomalies that could be present in the system.}.
The main distinction in the anomaly polynomials between ordinary Weyl fermions and multi-Weyl ones is that in the latter case the anomaly coefficients are enhanced by the topological charge $n$ \cite{LeporiLucaandBurrello2018AxialSemimetals,Dantas2020Non-AbelianSemimetals}. Consequently, the chiral magnetic and vortical effects manifest in the transport properties of mWSMs \cite{Dantas2018MagnetotransportApproach,Dantas2020Non-AbelianSemimetals}. 
In fact, the fascinating properties of topological semimetals have also garnered the attention of high-energy theorists who have explored their properties in a strongly coupled regime via the gauge-gravity duality \cite{Gursoy2012HolographicSemimetals, Landsteiner2016TheSemi-metal, Landsteiner:2016stv, Grignani:2016wyz, Juricic2024HolographicDislocations, Liu2021AnSemimetal, Rodgers2021ThermodynamicsSemimetals}.

The last decade has witnessed a reformulation of the hydrodynamic paradigm \cite{KovtunLecturesTheories} in terms of an action principle \cite{Haehl2015AdiabaticDissipation, Liu2017LecturesHydrodynamics}. The modern description of hydrodynamics is having a strong impact on our understanding of out-of-equilibrium systems, as it allows the systematic incorporation of hydrodynamic fluctuations and reveals intriguing connections with quantum chaos \cite{Grozdanov2019OnCorrections}. As a first approach to a hydrodynamic system, the partition function method \cite{Banerjee2012ConstraintsFunctions, Jensen2014AnomalyEquilibrium} is a robust technique that accounts only for non-dissipative effects; capturing the properties of ideal hydrodynamics. The Schwinger-Keldish effective action for dissipative systems and the partition function methods for ideal fluids have been successfully applied to a large variety of systems, such as relativistic spin-hydrodynamics \cite{Gallegos:2021bzp,Hongo:2021ona,Gallegos:2022jow}, Galilean fluids \cite{Rangamani:2008gi,Jensen2015AspectsTheory,Jain:2020vgc,Hartong:2024lqb}, fracton fluids \cite{Glodkowski:2022xje,Jain2023DipoleHydrodynamics,Armas:2023ouk,Glodkowski:2024ova,Jain2024DipoleII}, non-boost invariant fluids \cite{Armas2021EffectiveBoosts,Pinzani-Fokeeva:2021klb}, and they have served a pivotal role in the understanding of gauge and gravitational anomalies in the transport properties of chiral fluids \cite{Jensen2014AnomalyEquilibrium, Jensen2012ThermodynamicsCones}. The partition function method relies on the idea of interpreting the generating function of quantum field theory as the effective action for the hydrostatic regime of the finite density and temperature fluid phase of the theory. On one hand, diffeomorphism and gauge invariance fix the form of the Ward identities of the conserved currents, and on the other hand, they strongly constrain how the geometric and gauge fields enter in the effective action, once a derivative expansion has been fixed. 

Since spacetime-related conserved currents are sourced by deformations of the underlying spacetime geometry, it is mandatory to couple the microscopic system under consideration to a generic curved manifold $\mathcal{M}_{4}$ realizing the spacetime isometries group of the system when taken to be flat. The most common examples are pseudo-Riemannian manifolds as a natural background where Lorentzian fields propagate, Newton-Cartan spacetimes for Galilean field theories, and Aristotelian geometries for theories lacking boost symmetry \cite{Bacry:1968zf,Figueroa-OFarrill:2018ilb}. Nonetheless, the spacetime group $\mathcal G $ for the class of systems we discuss here is
\begin{equation}
\mathcal G = (SO(1,1)\times SO(2))\ltimes \mathbb R^4\,,
\end{equation}
implying that we must find the appropriate manifold $\mathcal{M}_{4}$, since the system is Lorentzian along a plane and Euclidean on the complementary space. Without loss of generality, we can take the coordinates $x^\mu=(x^a,x^i)$ with $a=0,1$ and $i=2,3$ such that $x^a$ expand the Lorentzian directions and $x^i$ the Euclidean ones. The desired set of external fields is then obtained by gauging the group $\mathcal G$ together with an internal $U(1)$.

In this paper, our goal is to extract properties of hydrodynamics from a covariant effective action. This constitutes two ingredients: 1) to find the constitutive relations, and 2) to extract the Ward identities arising from the symmetries.  Such a covariant effective action allows us to compute constitutive relations of the conserved currents, obtained after taking variations of the action with respect to the external fields that couple to them. On the other hand, conservation laws are obtained from the gauge invariance of the action.

The advantages of our general approach are that even if our motivation is the ideal flow of mWSMs in the hydrodynamic regime, our results are applicable to any system (fermionic or bosonic) with the space-time group $\mathcal G$. In fact, we point out that Bjorken fluids \cite{PhysRevD.27.140, PhysRevLett.130.241601} are characterized by the group  $\mathcal G_{\mathrm{Bjorken}} = SO(1,1)\times ISO(2)$, meaning that there is a class of the spacetimes we consider here that will have  $\mathcal G_{\mathrm{Bjorken}}$ as the isometry group \footnote{Bjorken flow is expected to capture the ultrarelativistic effects in heavy ion collision with the system required to be boost invariant along the direction defined by the beam $x^1$, and rotational and translational invariant in the $x^i$  plane. }. Therefore, our theory paves the road to the characterization and modeling of Bjorken fluids. Our main results consist of a covariant action for mWSMs compatible with the general coordinates principle, curved spacetime, and gauge fields, the Ward identities associated with the spacetime and internal charges of the problem, and the zeroth-order constitutive relations for the currents.

The paper is organized as follows: in Section~\ref{sec:Thermal_Equilibrium} we review general properties of the grand canonical ensemble for Lorentzian and Galilean systems and compare them with the thermal ensemble needed to describe equilibrium states for theories with the spacetime group $\mathcal{G}$. The peculiarities of a system with reduced boost symmetry are discussed, and the thermodynamic variables are determined. In Section~\ref{sec:multi-Weyl} we review the main properties of Weyl fermions and derive the low-energy action for noninteracting mWSMs from the theory proposed in \cite{Dantas2020Non-AbelianSemimetals}. Section~\ref{sect:CurvedSpace} is dedicated to obtaining the spacetime in which multi-Weyl particles propagate. After such a construction, a curved spacetime action is proposed, and the properties of the generating function are analyzed. Then, we constrain the form of the lowest-order partition function and use it to derive the ideal constitutive relations for the conserved currents in Section~\ref{sect:mW_PartitionFunction}. We conclude in Section~\ref{sect:discussion} with some closing remarks and outlooks.

\section{Thermal equilibrium vs boost symmetry}\label{sec:Thermal_Equilibrium}

In thermodynamics, the entropy $S(E,V,\mathbf Q)$ quantifies the number of microscopic states available to a closed system, and it depends on the internal energy ($E$), volume ($V$), and any additional conserved charges $\mathbf Q=(Q_1,\ldots,Q_N)$. Since we aim to investigate thermodynamic equilibrium and the small perturbations around it, we will focus on scenarios where $V\to\infty$ while keeping the densities $(\mathcal{E},\boldsymbol\rho)=(V^{-1}E,V^{-1}\mathbf Q)$ finite. Implicitly, when discussing thermodynamics, the role of boost invariance is assumed; therefore, most of our intuition is applicable to systems with boost invariance. 

To explore the significance of boost symmetry in thermodynamics, let us consider a system possessing a continuous symmetry group $\mathcal G$ accounting for both spacetime and internal transformations. Therefore, the Casimirs $\hat Q^I$ of the Lie algebra provide a set of invariants that define the system's state \footnote{In our notation, hatted variables denote the microscopic charges. Our discussion is general and applies both to classical or quantum systems.}. Consequently, the entropy will depend on their expectation values $Q^I=\langle\hat{Q}^I\rangle$ at the thermal state. For instance, in a Lorentzian system with a $U(1)$ internal charge the relevant charges are $\hat Q^I=(\hat E,\hat Q)$, with the energy given by the quadratic Casimir of the Poincar\'e group
\begin{equation}
\hat E^2=- \eta^{\m\nu}\hat P_\mu  \hat P_\nu\,,
\end{equation}
and $\eta^{\m\nu}=\mathrm{diag}(-1,1,1,1)$. 
On the other hand, if we consider a Galilean system, the charges are $\hat Q^I=(\hat E,\hat M)$
\begin{equation}
\hat M\hat E =  -\hat M \hat P_0 + \frac{1}{2} | \boldsymbol{\hat P}|^2\,, 
\end{equation}
with $\hat M$ and $\hat E$ denoting the central charge and the quadratic Casimirs of the Bargmann algebra, respectively. In both scenarios, $\hat P_0$ is the generator of time translations and the internal energy $E$ is the negative of the expectation value of $\hat P_0$ at the rest-frame of the system.

Conversely, for a system characterized by the group $\mathcal G=(SO(1,1)\times SO(2))\ltimes \mathbb R^4$ the quadratic Casimirs are
\begin{equation}\label{Casimirs}
\hat E^2= -\eta^{ab}\hat P_a \hat P_b\,, \qquad \hat{ \mathbb P}^2=  (\hat P_i)^2 \,,
\end{equation}
where indices $a,b=0,1$ and $i,j=2,3$, and $\eta^{ab}={\rm diag}(-1,1)$ represents the two-dimensional Minkowski metric. Without loss of generality, we can assume that $P_\mu = ( -E,0,0,\mathbb P)$ and after introducing a $U(1)$ charge $Q$ the first law of thermodynamics reads 
\begin{equation}
dE = TdS + \a\, d \mathbb P + \mu dQ -\mathcal PdV\,,
\end{equation}
with $\mathcal P$ the pressure and $\a,\mu$ the respective chemical potentials, canonical to the charges $P_3=\mathbb P$, and $Q$. 
 Then, we  introduce the entropy $s=S/V$, momentum $ p = \mathbb P/V$ and $U(1)$ $\rho=Q/V$ densities together with the Gibbs-Duhem relation 
\begin{align}
\mathcal E+\mathcal  P &=T s+\a\, p +\mu\rho\,,
\end{align}
that allows to write the differential of the pressure as follows 
\begin{align}
d\mathcal P &= s dT + p d\a + \rho d\mu \,,
\end{align}
implying that pressure is a function of the temperature and chemical potentials, i.e. $\mathcal{P}=\mathcal P\left(T,\a,\m \right)$. 

In what follows, we describe this class of theories by introducing the grand canonical ensemble. This approach allows for a reformulation of the theory within the framework of quantum statistical mechanics using the path-integral method, thereby extending the thermodynamic domain to the so-called hydrostatic regime. Thus, we define the partition function as \(\mathcal{Z}_0(T,\a,\mu) = \mathrm{Tr}\varrho_0(\b,\bar\a,\bar\mu)\), where the set of variables \((\b,\bar\a,\bar\mu)\) denotes the inverse temperature, \(\b = T^{-1}\) and the `reduced' chemical potentials, $(\bar\a,\bar\mu)=\b(\a,\m)$, respectively. It will function, as is typical, as a generating function to derive the expectation values of the densities. Moreover, the density operator is defined as follows
\begin{equation}
\varrho_0(\b,\bar\a,\bar\mu) =  e^{\b\hat P_0 +\bar\a \hat P_3 +\bar \mu\hat Q}\,.
\end{equation}
Notice that, by acting with a generic symmetry transformation $\mathcal{B}(u^a,v^i)\in SO(1,1)\times SO(2)$ with parameters satisfying $u_au^a=-1$ and $(v_i)^2=1$, 
\begin{equation}
\mathcal{B}(u^a,v^i)\hat P_\mu\mathcal{B}^{-1}(u^a,v^i)\to \Lambda_\mu\,^\nu (u^a,v^i)\hat P_\nu\,,
\end{equation}
the density operator can be boosted/rotated to a two-parameter family of `inertial' frames, for which the density $\varrho$ reads
 \begin{equation}
\varrho(\beta^\mu,\bar \mu) = \mathcal{B}(u^a,v^i) \varrho_0(\b,\bar\a,\bar\mu)\mathcal{B}^{-1}(u^a,v^i) = e^{\beta^\mu\hat P_\mu  + \bar \mu\hat Q}\,,
\end{equation}
where the `thermal velocity'  is $\beta^\mu=(\beta u^a,\bar \a v^i)$. Thus, we deduce that for this family of theories, their partition function agrees with that of a system with internal energy $E=-\langle \hat P_0\rangle$ and momentum $k=\langle \hat P_3\rangle$, i.e.
 \begin{equation}
\mathcal  Z(\beta^\mu ,\mu) = \mathcal  Z_0(T,\a,\mu)\,.
 \end{equation}
This allows us to interpret the chemical potential $\a$ as the magnitude of the relative velocity between the observer and the system. Consequently, we define the velocity of the bath as $U^\mu = T\beta^\mu$.
On the other hand, the partition function can be expressed as a Euclidean effective action after using the fact that the grand canonical potential  $\Omega =-T\log\mathcal Z$   is related with the pressure via
 \begin{equation}
 \label{eq_Thermal_Partion_Funcion_1}
\log\mathcal Z=  \b V\,\mathcal P \,.
 \end{equation}
Interestingly, we notice that the thermodynamic potential $\tilde{\mathcal E} =\mathcal  E - \a \kappa$ (see \cite{DeBoer2018PerfectFluids}) satisfying 
\begin{align}
d\tilde{\mathcal E} &= T ds  + \mu d \rho - p d\a\,,\\
  \tilde{\mathcal E} + \mathcal  P &= T s + \mu \rho  \,,\label{eq_Gibbs-Duhem2}
\end{align}
can be interpreted as the internal energy measured by a `non-inertial' observer, at rest relative to the thermal bath.%, using  $\hat{\tilde{P}}_0 = U^\mu\hat P_\mu$ as the generator of time translations. 

\section{Symmetries and action principle for multi-Weyl semimetals}\label{sec:multi-Weyl}

Multi-Weyl semimetals represent a class of topological materials characterized by the presence of gapless quasiparticles that exhibit Weyl-like behavior near the Fermi level. In some sense, for Multi-Weyl fermions, the concept of chirality is extended by attributing an integer topological (Berry) charge $\pm n$, where $n> 1$. Remarkably, the Berry monopole in Weyl semimetals shows a tight connection with the celebrated chiral anomalies in quantum field theory \cite{Dantas2020Non-AbelianSemimetals,Dantas2018MagnetotransportApproach,LeporiLucaandBurrello2018AxialSemimetals}. Nevertheless, contrary to ordinary Weyl (Dirac) fermions, multi-Weyl particles are not Poincar\'e invariant.

Before delving into the properties of this class of exotic fields, we find it convenient to review some of the main properties of ordinary Weyl fermions. To do so, we introduce a dual basis of hermitian matrices $(\sigma^\mu,\bar\sigma^\mu)$ 
\begin{equation}
\sigma^\mu =(\mathbb I,\boldsymbol \sigma)\,,\qquad \bar \sigma^\mu =(\mathbb I,-\boldsymbol{\sigma})\,,
\end{equation}
satisfying ${\rm Tr}[\bar\sigma^\mu\sigma^\nu]=-2\eta^{\mu\nu}$. Therefore, the right- and left-handed Weyl operators are \footnote{Notice that after a parity transformation $(p_0,\boldsymbol p)\to (p_0,-\boldsymbol p)$ the left and right operators are exchanged.}
\begin{equation}\label{eq_Weyl_op}
\mathcal D^{(R)} = \sigma^\mu p_\mu\,,\qquad \mathcal D^{(L)} = \bar\sigma^\mu p_\mu\,,
\end{equation}
where $p_\mu=(-E,\mathbf p)$.  Since $\mathrm{Det}\, \mathcal D^{(L,R)}=-p_\mu p^\mu$ and $\mathrm{Tr}\, \mathcal D^{(L,R)}=2p_0$ the realization of the Lorentz group on Weyl fermions is in terms of $SL(2,\mathbb C)$ matrices. In fact, given a $SL(2,\mathbb C)$ element $S$ the $\sigma^\mu$ matrices obey the relations
\begin{equation}
\sigma^\mu(\Lambda^{-1})^\nu\,_\mu = (S^{-1})^\dag\sigma^\nu S^{-1}\,,\qquad \bar\sigma^\mu (\Lambda^{-1})^\nu\,_\mu = S\sigma^\nu S^\dag\,,
\end{equation}
where $(\Lambda^{-1})^\nu\,_\mu$ is a Lorentz matrix. Therefore, the action of the Lorentz group on the Weyl operators reads 
\begin{equation}\label{eq_TransformWeyl_op}
\mathcal D'^{(R)}|_{p'} = (S^{-1})^\dag\mathcal D^{(R)}|_{p} S^{-1}\,,\qquad \mathcal D'^{(L)}|_{p'} = S\mathcal D^{(L)}|_{p}S^\dag\,,
\end{equation}
implying that  right- and left-handed fermions $\Psi_R,\Psi_L$ obey the transformation rule $\Psi_R\to S\Psi_R$$, \Psi_L\to (S^{-1})^\dag\Psi_L$. As a consequence of previous properties, the Weyl equations are covariant under the action of the Lorentz group, and $p_\m$ has to be null for on-shell particles.

To introduce the multi-Weyl case, we start from the model proposed in \cite{Dantas2020Non-AbelianSemimetals} by defining a set of Weyl fermions with flavor symmetry $SU(2)_n$, transforming in the spin $j=(n-1)/2$ representation of the group.  A system of Weyl fermions with such a symmetry group exhibits multi-Weyl physics when coupled to a constant background field $A = \frac{m}{2}\delta_{\bar k k}\tau^{\bar k}\,dx^k$, with $\bar k,k=1,2$ and $(\tau^{\bar k},\tau^3)$ generators of the internal flavor $su(2)_n$ Lie algebra. The right-handed Weyl operator for this system is modify as $\mathcal{D} = p_\mu\sigma^\mu + \hat{\mathcal{ D}}_n$ with 
\begin{equation}\label{eq_flavorMWeyl}
\hat{\mathcal{D}}_n = \frac{m}{2}\delta_{k\bar k} \sigma^k \tau^{\bar k}\,,
\end{equation}
where $m$ is a parameter with the dimension of mass, and for clarity in the notation we have removed the label $(R)$ in the Weyl operator. The left-handed operator is obtained after sending $\mathbf p\to -\mathbf p$. 
In fact, the $\hat{\mathcal{ D}}_n$ term breaks the $SL(2,\mathbb C)\times SU(2)_n$ group down to the diagonal subgroup $ SO(1,1)\times U(1)$, where $SO(1,1)$ refers to boosts in the $x^3$ direction, whereas the $U(1)$ represents a diagonal rotation around the third spacetime and internal directions\footnote{In  this section, contrary to the rest of the paper, but without lost of generality, we label the Euclidean coordinates $x^i=(x^1,x^2)$, whereas the Lorentzian one are $x^a=(x^0,x^3)$.}. The set of unbroken symmetry transformations reads
\begin{equation}
T(\lambda^{(S)},\lambda^{(L)}) = \exp\left[\frac{1}{2}\lambda^{(S)}\sigma^3 + i\lambda^{(L)}\left(\frac{1}{2}\sigma^3 + \tau^3\right)\right]\,.
\end{equation}
 These transformations, in addition to leaving the four-dimensional Minkowski metric and the Levi-Civita tensor invariant, preserve also the background field strength $F=m^2\tau^3dx^1\wedge dx^2 $. This induces a degenerate area element on the $SO(2)$ plane, which we will call $\varepsilon$. Therefore, the set of independent tensors is given by
\begin{equation}
G=\eta_{\mu\nu}dx^\mu dx^\nu\,,\quad dV=\frac{1}{4!} \epsilon_{\mu\nu\rho\lambda}dx^\mu\wedge dx^\nu\wedge dx^\rho\wedge dx^\lambda\,,\quad \varepsilon=\frac{1}{2m^2}F^3_{\mu\nu}dx^\mu\wedge dx^\nu\,.
\end{equation}

The background magnetic field $F$ gaps-out a sector of the theory, while keeping one massless mode. At sufficient low energies, this mode behaves as a multi-Weyl particle. To identify the gapless sector of the system, let us begin by setting $p^i=0$ in the Weyl operator   
\begin{equation}\label{eq_flavorMWeyl2}
\mathcal{D} = p_a\sigma^a + m ( \sigma^+ \tau^- + \sigma^- \tau^+)\,,
\end{equation}
and note that we have introduced the matrices $\sigma^\pm= \frac{1}{2}(\sigma^1\pm i\sigma^2)$ and $\tau^\pm= \frac{1}{2}(\tau^1\pm i\tau^2)$. In addition,  we introduce the basis $|\pm,m_j\rangle=|1/2,\pm\rangle\otimes|j,m_j\rangle$ with $m_j = -j,\ldots ,j$ which satisfy $\sigma^3|\pm,m_j\rangle = \pm|\pm,m_j\rangle$, $\tau^3|\pm,m_j\rangle=m_j|\pm,m_j\rangle$, and by inspection, we conclude that the pair of vectors 
\begin{equation}
\phi^0_\pm  = |\pm,\pm (n-1)/2 \rangle\,.
\end{equation}
are eigenvectors of  Eq.~\eqref{eq_flavorMWeyl2} with eigenvalues $p_0\pm p_3$. Therefore, they expand the subspace of massless particles at $p^i=0$. Interestingly, we note the tensorial nature of the gapless subspace since $T(\lambda^{(L)})\phi^0_\pm = e^{\pm i\, n \lambda^{(L)} /2}\phi^0_\pm$ .
  On the other hand, the symmetric (antisymmetric) combinations
\begin{align}
&( |+,n/2-3/2 \rangle\, \pm\, |-,n/2-1/2 \rangle )\,, 
\ldots \,,
( |+,1/2-n/2 \rangle\, \pm\, |-,3/2-n/2 \rangle)\,,
\end{align}
 expand the gapped sector of positive (negative) energy modes. 
At low enough energies $|\mathbf p |/m \ll 1$, the massive fields can be integrated out and as we anticipated above, the effective theory 
is that of multi-Weyl fermions (for a detailed derivation see \cite{Dantas2020Non-AbelianSemimetals}) and the (multi-)Weyl operator reads
\begin{equation}\label{eq_MultiWeyl_Op}
\mathcal D_{eff} =  p_a\sigma^a + m^{1-n}(p_+^n\sigma^- +  p_-^n\sigma^+)\,,
\end{equation}
where the complex momentum is $p_+ = p_1+ip_2$.
Finally, after Fourier transforming and using the properties of the Pauli matrices, we can write the action for a right-handed multi-Weyl fermion as
\begin{equation}\label{eq_Flat_Action}
    S = \int \, \Re\,\bar\phi\left[i \slashed\6_\parallel + \,m^{1-n} (i\slashed\6_\perp)^n \right]\phi\,dV \,.
    \end{equation}
where $\bar\phi=\phi^\dag\gamma^0$, $\slashed\6=\gamma^\mu\6_\mu$ and the gamma matrices are $\gamma^\mu=(\sigma^1,\sigma^0,-i\sigma^3, -i\sigma^2)$. Notice that we have defined the longitudinal and transverse gamma matrices by contracting them with the projectors $g^\mu_\nu=\delta^\mu_\nu-h^\mu_\nu$ and $h^\mu_\nu=\varepsilon^{\mu\rho}\varepsilon_{\rho\nu}$, respectively.

Since our goal is constructing a covariant partition function for this class of systems, in the next section, we shall proceed to gauge the global symmetries of the system. This procedure, allows one compute the one-point functions of the conserved operators. As we are working with fermionic fields, it is not enough to consider a generic curved background, but a frame bundle compatible with the (reduced) boost and rotational invariance of the system. 
More precisely, it is necessary to construct first a frame bundle with the (flat space) isotropy group as the structure group of the bundle. Then, adding a spin structure is natural, since the spin group is the double cover of the orthogonal group. For ordinary relativistic fermions, the spacetime must have the structure of a $SO(d,1)$ principal bundle, however, the class of theories we discuss here are not invariant under the full Lorentz group, but under the subgroup $SO(1,1)\times SO(2)$. Therefore, to find the appropriate curved geometry where the fermions are defined, we need to gauge the space-time symmetries.

\section{Multi-Weyl particles in curved spacetime}\label{sect:CurvedSpace}

The computation of correlation functions of conserved charges in quantum field theory can be accessed via the so-called generating functional or effective action of the system. Such a functional is obtained after integrating out the matter content that has been minimally coupled to external background fields. In particular, the expectation value of conserved currents associated with spacetime symmetries is obtained after coupling the system to a curved background. Generically, Lorentzian fields are coupled to pseudo-Riemannian geometries, Galilean particles to Newton-Cartan spaces, whereas theories lacking boost invariance need to be coupled to the so-called Aristotelian spacetime \cite{Bacry:1968zf,Bidussi:2021nmp,Armas2021EffectiveBoosts}. However, the class of systems we are interested in does not fit within these examples, and the appropriate geometry compatible with the reduced boost and rotational symmetry of the problem needs to be constructed. To do so, we will gauge the spacetime and internal symmetry group of the problem under consideration.

\subsection{External background geometry as a $(SO(1,1)\times SO(2))\ltimes \mathbb R^4\times U(1)$ gauge theory}

The required external fields can be encoded in a gauge connection $\mathcal A$ taking values on the Lie algebra associated to the group $\mathcal G=(SO(1,1)\times SO(2))\ltimes \mathbb R^4\times U(1)$. In addition to the invariant tensors of the $3+1$ Poincar\'e group $\eta_{AB},\epsilon_{ABCD}$ with $A=0,1,2,3$, $\mathcal G $ possesses invariant $SO(2)$ projector  $h^A_B$ with trace $h=2$, and the degenerate antisymmetric tensors $\epsilon_{AB}$, $\varepsilon_{AB}$ satisfying
\begin{equation}
h^A_A=2\,, \qquad h^A_Bh^B_C=h^A_C\,,\qquad h^A_B\epsilon_{AC}=0\,,\qquad h^A_B\varepsilon_{AC}=\varepsilon_{BC}\,,
\end{equation}
using the Minkowski metric to lower and raise indices we can define the  Euclidean $h_{AB}$ and Lorentzian $g_{AB}=\eta_{AB}-h_{AB}$ `metrics'. Unless explicitly specified we will always use the Minkowski metric to raise and lower tangent indices. The antisymmetric tensors satisfy also
\begin{equation}
\varepsilon^{AB}\varepsilon_{CD}=2h^{[A}_C h^{B]}_D\,, \qquad \epsilon^{AB}\epsilon_{CD}=2g^{[B}_C g^{A]}_D\,,
\end{equation}
where $g^A_B$ is the $SO(1,1)$ projector
\begin{equation}
g^A_B=\delta^A_B-h^A_B, \qquad  g^A_B g^B_C=g^A_C \,.
\end{equation}
Using the properties of the aforementioned tensors, we represent the elements of the Lie algebra $\mathfrak so(1,1)\oplus\mathfrak so(2)$ in the form
\begin{equation}
M_{AB}=-S\,\epsilon_{AB} 
+ L\,\varepsilon_{AB}\,.
\end{equation}
On the other hand we denote the generators of spacetime translations and $U(1)$ transformations $P_A,Q$  respectively.
Therefore the non-vanishing Lie brackets are
\begin{equation}
[M_{AB},P_C] = (\epsilon_{AB}\epsilon_C\,^D -\varepsilon_{AB}\varepsilon_C\,^D)P_D\,,
\end{equation}
and the Lie algebra valued one-form gauge connection is defined as
\begin{equation}\label{ConnectionOne-form}
\mathcal A= e^A P_A  + \frac{1}{2}\Omega^{AB} \,M_{AB}  + A \,Q\,.
\end{equation}
where $A$ is an Abelian $\mathfrak u(1)$ electromagnetic potential, $e^A$ is interpreted as the vierbeins, and $\Omega^{AB}$ is defined as \footnote{In a system permitting a spontaneous breaking of dilatation symmetry in the presence of a U$(1)$ gauge field, two gauge fields are required which is being studied in \cite{gapped-dilaton}. }
\begin{equation}\label{OmegaABdef}
\Omega^{AB}= \Omega \,\epsilon^{AB}+ \omega\, \varepsilon^{AB}\,,
\end{equation}
in terms of Abelian spin connections $\Omega$ and $\omega$ , associated to the Lie algebras $\mathfrak{so}(1,1)$ and $\mathfrak{so}(2)$, respectively. The curvature two-form associated with the connection \eqref{ConnectionOne-form} reads
\begin{equation}\label{CurvatureTwo-form}
\mathcal F=d\mathcal A+ \mathcal A\wedge\mathcal A= T^A P_A  + \frac{1}{2}\mathcal F^{AB}\, M_{AB} + F\,Q
\end{equation}
where we have defined curvatures 
\begin{equation}
\mathcal F^{AB}= \mathcal F^{(S)} \,\epsilon^{AB}+ \mathcal F^{(L)}\, \varepsilon^{AB}=d\Omega^{AB}\,,\qquad F = dA\,,
\end{equation}
and the torsion two-form
\begin{equation}\label{torsions}
T^A=De^A\,,
\end{equation}
with the exterior covariant derivative defined as 
\begin{equation}
D= d + A\,\mathcal P[Q] + \frac{1}{2}\Omega^{AB}\, \mathcal P[M_{AB}] \,,
\end{equation}
with $\mathcal P[Q]$ and $\mathcal P[M_{AB}]$ being the corresponding representations for the charge, boost and angular momentum generators. In particular, it acts on vector- and fermion-valued fields as follows
\begin{subequations}\label{Dvectorfermion}
\begin{align}
&DV^A = dV^A  + \Omega^A\,_B  V^B\,, \\
&D\Psi = \left[d + iA +  \frac{1}{2} (\Omega 
+ i\,n\,\omega)\tau^3\right]\Psi\,.
\end{align}
\end{subequations}
In addition, we introduce inverse vierbeins $E_A$, satisfying the orthogonality relation
\begin{equation}\label{OR}
\begin{aligned}
&e^A{}_\mu E_B{}^\mu = \delta^A_B,  
&\qquad&
e^A{}_\mu E_A{}^\nu=\delta^\nu_\mu\,,
\end{aligned}
\end{equation}
and the flat geometry of the previous section is recovered after setting $e^A=dx^A$, $\Omega^{AB}=A=0$. Since the vierbeins are invertible we can introduce the volume form 
\begin{equation}
dV=e^0\wedge e^1\wedge e^2\wedge e^3=|e|d^4x\,.
\end{equation}
Using $\epsilon_{AB}$ and $\varepsilon_{AB}$ we can define the two forms
\begin{align}
    \epsilon =\frac{1}{2}\epsilon_{AB}e^A{}\wedge e^B{}\,, \qquad \varepsilon =\frac{1}{2}\varepsilon_{AB}e^A{}\wedge e^B{}\,,
\end{align}
and express the volume form as $dV=\epsilon\wedge\varepsilon$. On the other hand the three dimensional area element reads
\begin{equation}\label{defdSA}
(dS)_A = \epsilon_A\wedge\varepsilon -\varepsilon_A\wedge\epsilon\,,
\end{equation}
with $\epsilon_A=\epsilon_{AB}\,e^B$ and $\varepsilon_A=\varepsilon_{AB}\,e^B$.

Notice that the vierbeins and their inverse can be used to trade tensors into Lie algebra valued tensors and viceversa, in particular, using $\eta_{AB}$, $g^A_B$, and $h^A_B$ we can defined the set of tensors
\begin{align}
    g^\mu_\nu = g^A_BE_A\,^\m e^B\,_\nu\,, \quad h^\mu_\nu = h^A_BE_A\,^\m e^B\,_\nu\,, \quad
    G_{\mu\nu} =  e^A{}\,_\mu e_{A\nu} \,, \quad
    G^{\mu\nu} = E^{A\mu} E_A\,^\nu,
\end{align}
satisfying
\begin{align}
    g^\mu_\nu + h^\mu_\nu = \delta^\mu_\nu\,, \quad
    G^{\mu\beta}G_{\beta\nu}  =  \delta^\mu_\nu\,,
\end{align}
similarly to the case of tangent space indices we will use $G^{\m\n}$ and $G_{\m\n}$ to raise and lower curved space indices.

\subsection{Gauge transformations}
The transformation law of the connection under infinitesimal diffeomorphisms and gauge transformations with parameters $\chi=(\xi,\Lambda)$, where 
\begin{equation}\label{gaugepar}
\Lambda= \frac{1}{2}\lambda^{AB} M_{AB} + \lambda^{(Q)} Q
\end{equation}
and $\lambda^{AB} = \epsilon^{AB} \lambda^{(S)}+ \varepsilon^{AB} \lambda^{(L)}$, reads
\begin{equation}
\delta_\chi \mathcal A =\mathfrak L_\xi \mathcal A + \mathcal D\Lambda
\end{equation}
where $\mathcal D= d+ [\mathcal A,\;]$. 
Evaluating this transformation for each field, we find
\begin{subequations}\label{Tgf1}
\begin{align}
& \delta_\chi e^A = \mathfrak L_\xi e^A  - \lambda^{A}\,_B\,e^B= D\xi^A + \bar{\mathcal{ G}}^A - \bar\lambda^{A}\,_B e^B{}  \,, 
\\
& \delta_\chi \Omega^A\,_B = \mathfrak L_\xi \Omega^A\,_B + d\lambda^{A}\,_B = d \bar\lambda^{A}\,_B + \bar{\mathcal{ E}}^{A}\,_B\,,
\\
& \delta_\chi A = \mathfrak L_\xi A+ d\lambda^{(Q)}  = d \bar\lambda^{(Q)}+ \bar{ E}\,,
\end{align}
\end{subequations}
where we introduced new parameters
\begin{equation}\label{eq_covGaugeParam}
\begin{aligned}
\bar\lambda^{AB} & = \lambda^{AB} + \iota_\xi \Omega^{AB}\,,\qquad
\bar\lambda^{(Q)}  = \lambda^{(Q)} + \iota_\xi A\,,
\end{aligned}
\end{equation}
and the ``electric'' fields
\begin{equation}
\begin{aligned}
\bar{\mathcal{ G}}^A = \iota_\xi T^A\,,\qquad \bar{\mathcal{ E}}^{AB} & = \iota_\xi \mathcal F^{AB}= \iota_\xi \mathcal F^{(S)} \,\epsilon^{AB}+ \iota_\xi \mathcal F^{(L)}\, \varepsilon^{AB}\,,\qquad
\bar{ E}  = \iota_\xi  F\,.
\end{aligned}
\end{equation}
These relations can be easily obtained by applying Cartan's magic formula $\mathfrak L_\xi \mathscr{F}= \iota_\xi d\mathscr{F}+d \iota_\xi\mathscr{F}$, where $\mathscr{F}$ is a differential form. \\
It is convenient to extend the individual Lie algebras of diffeomorphisms and gauge transformations by introducing the bracket 
\begin{equation}
[\chi',\chi]=\delta_{\chi'}\chi = (\mathfrak L_{\xi'}\xi, \mathfrak L_{\xi'}\Lambda - \mathfrak L_{\xi}\Lambda')\,,
\end{equation}
after some tedious derivations it is possible to prove that
\begin{equation}
[\delta_{\chi'},\delta_\chi] =\delta_{[\chi',\chi]}\,,
\end{equation}
and that $\bar\lambda^{(S)}$, $\bar\lambda^{(L)}$, $\bar\lambda^{(Q)}$ transform as scalar functions.

\subsection{Intrinsic and extrinsic torsion}
A spacetime with $SO(1,1)\times SO(2)$ structure group defines a generalized Aristotelian geometry \cite{Figueroa-OFarrill:2020gpr}  exhibiting a foliation analogous to the string version of Newton-Cartan geometry \cite{Andringa:2012uz,Deveciolu2019StringGeometry,Bidussi:2021ujm}. In contrast to standard Riemannian geometry, not all components of the torsion depend on the spin connection. This becomes manifest when splitting the torsion two-forms \eqref{torsions} in the following way
\begin{equation}\label{T+T}
T^A = \mathring T^A + \hat T^A
\end{equation}
where
\begin{equation}\label{IntTorComps}
\mathring T^a=\frac12   T^a{}_{ij}e^i\wedge e^j +  T^{(ab)}{}{}_i e_b \wedge e^i,
\qquad
\mathring{ T}^i= \frac12  T^i{}_{ab}e^a\wedge e^b + T^{(ij)}{}{}_ae_j \wedge e^a,
\end{equation}
and
\begin{equation}\label{ExtTorComps}
\hat T^a= \frac12  T^a{}_{bc} e^b\wedge  e^c + T^{[ab]}{}{}_i e_b \wedge e^i,\qquad
\hat{ T}^i= \frac12  T^i{}_{jk } e^i\wedge e^j + T^{[ij]}{}{}_a e_j\wedge e^a\,.
\end{equation}
From now on, without a loss of generality, we assume that $g^A_B=\mathrm{diag}(1,1,0,0)$ and $h^A_B=\mathrm{diag}(0,0,1,1)$ and we introduce indices $a=0,1$ and $i=2,3$ to distinguish between the longitudinal and transverse components of $SO(1,1)\times SO(2)$ tensors\footnote{Note that this definition must be adapted when applied to the results of Section~\ref{sec:multi-Weyl}, where the indices are split as $a=0,3$ and $i=1,2$.}.
The components of $\mathring T$ can be expressed solely in terms of the non-holonomic coefficients
\begin{equation}
\Sigma_{AB}{}{}^{C}=-2E_A{}^\mu E_B{}^\nu \partial_{[\mu}e^C{}_{\nu]},
\end{equation}
defined via the Lie brackets of the inverse vierbeins
\begin{align}
[E_A,E_A] = \Sigma_{AB}\,^C E_C\,.
\end{align}
This yields
\begin{equation}\label{TSigma}
\mathring T^a{}_{ij}=-\Sigma_{ij}{}{}^a 
\,,\quad
\mathring{ T}^i{}_{ab}=-\Sigma_{ab}{}{}^i   \,,\quad
\mathring T^{(ab)i}=\Sigma^{i(ab)} 
 \,,\quad
\mathring{ T}^{(ij)a}=\Sigma^{a(ij)}
 \,.
\end{equation}
Consequently, setting the components of $\mathring T^a$ and $\mathring T^i$ to zero does not contribute to expressing any spin connection components in terms of the tetrad and its derivatives. Instead, it imposes constraints on the geometry. To avoid such constraints and preserve the independence of all vierbein components, these torsion components will not be set to vanish and will be referred to as intrinsic torsion. In contrast, $\hat T^A$ depends on the spin connection components, which can be expressed as
\begin{subequations}
\begin{align}
&\hat T^a{}_{bc}=-\Sigma_{bc}{}{}^a
-2\epsilon^a{}_{[b}  E_{c]}{}^\mu\Omega_{\mu} \,,\qquad
\hat{ T}^i{}_{jk}=-\Sigma_{jk}{}{}^i
-2\varepsilon^i{}_{[j} E_{k]}{}^\mu\omega_{\mu} \,,
\\
&\hat T^{[ab]i}=-\Sigma^{i[ab]}- \epsilon^{ab} \Omega_{\mu}E^{i\mu}
\,,\qquad
\hat{ T}^{[ij]a}=-\Sigma^{a[ij]}-\varepsilon^{ij} 
 \omega_{\mu}  E^{a\mu}  
\,.
\end{align}
\end{subequations}
with $\epsilon_0\,^1=\epsilon_1\,^0=\varepsilon_2\,^3=-\varepsilon_3\,^2=1$. Thus, when $\hat T^A$ vanishes, $\Omega$ and $\omega$ can be expressed in terms of the vierbein. These components define the extrinsic torsion of the geometry, which may or may not be set to zero depending on whether the spin connections are treated as independent fields. \\
It is important to notice that the space-time manifold is not a product manifold of the form $\mathcal M_4= \mathcal M_{1,1}\times \mathcal M_2$, with $\mathcal M_{1,1}$ and $\mathcal M_2$ pseudo-Riemannian and Riemannian spaces, respectively. If it were, there would be no intrinsic torsion, and the number of independent fields ($\delta^a_\mu \delta^b_\nu g_{ab},\delta^i_\mu \delta^j_\nu h_{ij}$) would be $6$, not matching with the number of independent (conserved) spacetime-related operators of the theory ($T^\mu\,_\nu$), which is $10$. The proper way of sourcing all the operators requires a generic four dimensional manifold whose tangent space can be decomposed locally as $T_x\mathcal M_4 = \mathbb R^{1,1}\times\mathbb R^2$. This is the curved space-time we consider here. On a manifold with such a structure it is convenient to define a covariant derivative satisfying the property  $ \nabla_X Y_\parallel = P_\parallel(\nabla_X Y)$ and $ \nabla_X Y_\perp = P_\perp(\nabla_X Y)$, with $P_\parallel$, $P_\perp$ the projectors to the Lorentzian  and Euclidean slices of the tangent space respectively, i.e. $Y_\parallel=P_\parallel(Y)$, $Y_\perp=P_\perp(Y)$. If we recall that torsion is the failure of the connection to match the Lie bracket
 \begin{align}
 T(X,Y) &= \nabla_X Y - \nabla_Y X + [X,Y]\,,
 \end{align}
we can derive the following set of consistency conditions
 \begin{align}
\mathrm{if}\quad X=P_\parallel(X),\quad  Y=P_\parallel(Y) : \qquad P_\perp( T(X,Y) )&=  [X,Y]_\perp\,, \\
\mathrm{if}\quad X=P_\perp(X),\quad  Y=P_\perp(Y) : \qquad    P_\parallel(T(X,Y)) &= [X,Y]_\parallel\,,\\
\mathrm{if}\quad X=P_\parallel(X),\quad  Y=P_\perp(Y) : \qquad    P_\parallel(T(X,Y)) &= - \nabla_Y X +[X,Y]_\parallel\,,\\
P_\perp(T(X,Y)) &=  \nabla_X Y +[X,Y]_\perp\,.
 \end{align}
Using the basis \eqref{OR} satisfying,  $P_\parallel(E_a) = E_a$, $P_\parallel(E_i) = 0$, $P_\perp(E_i) = E_i$, $P_\perp(E_a) = 0$, we conclude that
\begin{align}
 T_{iab}&=  [E_a,E_b]_i\,, \\
T_{aij} &= [E_i,E_j]_a\,,\\
  T\,_{(ab)i} &= \frac{1}{2}( [E_b,E_i]_a + [E_a,E_i]_b)\,,\\
T\,_{(ij)a} &= \frac{1}{2}([E_j,E_a]_i + [E_i,E_a]_j)\,.
 \end{align}
where, in the last two lines, we have used $\nabla_X E_a = X^\mu\omega_\mu\epsilon_a\,^bE_b$ and $\nabla_X E_i = X^\mu\Omega_\mu\varepsilon_i\,^jE_j$. These are precisely the intrinsic torsion components \eqref{IntTorComps}. Thus, we conclude that, contrary to the Lorentzian/Riemannian case, the only components of the torsion field that we are free to set to zero are the ones given in \eqref{ExtTorComps}, namely,
 \begin{equation}
 T^\mu\,_{\nu\rho}e_{a\mu}E_b\,^\nu E_c\,^\rho  \,, \qquad T^\mu\,_{\nu\rho}e_{[a\mu}E_{b]}\,^\nu E_i\,^\rho \, \qquad T^\mu\,_{\nu\rho}e_{i\mu}E_j\,^\nu E_k\,^\rho  \,, \qquad T^\mu\,_{\nu\rho}e_{[i\mu}E_{j]}\,^\nu E_a\,^\rho\,.
 \end{equation}
Therefore, the Aristotelian geometry possesses a nonzero torsion. 

As is common in non-Lorentzian geometry \cite{Figueroa-OFarrill:2020gpr,Bidussi:2021nmp,Bergshoeff:2023rkk,Hartong:2024hvs}, the components of the intrinsic torsion can be expressed in terms of the so-called ``extrinsic curvatures''
\begin{equation}\label{extcurvatures}
K^a{}_{\mu\nu}=-\frac 12 \mathcal L_{E^a} h_{\mu\nu}, \qquad 
 K^i{}_{\mu\nu}=-\frac 12 \mathcal L_{E^i} g_{\mu\nu}\,,
\end{equation}
satisfying the following properties
\begin{subequations}
\begin{align}
&K^{(a}{}_{\mu\nu} E^{b)\nu}=0= K^{(i}{}_{\mu\nu} E^{j)\nu}\,,
\\
&K^{abc}=K^a{}_{\mu\nu} E^{b\mu} E^{c\nu}=0\,,
\\
& K^{ijk}= K^i{}_{\mu\nu} E^{j\mu} E^{k\nu}=0\,,
\end{align}
\end{subequations}
where the first relation implies, in particular, the identities
\begin{equation}\label{EinvKid}
K^a{}_{\mu\nu}E_a{}^\nu= K^i{}_{\mu\nu}E_i{}^\nu = 0\,.
\end{equation}
In terms of Eqs.~\eqref{extcurvatures}, the intrinsic torsion can be written as
\begin{equation}\label{TKs}
\mathring T^a= K_{ij}{}{}^a e^i \wedge e^j
+ K^{iab}e_b \wedge e_i\,, \qquad
\mathring T^i= K_{ab}{}{}^i e^a \wedge e^b
+ K^{aij}e_j \wedge e_a\,.
\end{equation}
Using the relations \eqref{extcurvatures} it is possible to understand $K^a{}_{\mu\nu}$ and $K^i{}_{\mu\nu}$ as the projections of the following object
\begin{equation}\label{Krhomunu}
\begin{aligned}
K^\rho{}_{\mu\nu} &= E_a{}^\rho K^a{}_{\mu\nu}+ E_i{}^\rho K^i{}_{\mu\nu}
\\
&=  \frac 12g^{\rho\lambda} \left( 2 \partial_{(\mu} h_{\nu)\lambda}-\partial_\lambda h_{\mu\nu}\right)
+\frac 12 h^{\rho\lambda}\left( 2 \partial_{(\mu} g_{\nu)\lambda}- \partial_\lambda g_{\mu\nu} \right)\,.
\end{aligned}
\end{equation}
Note that the identities \eqref{EinvKid} imply that $K^\rho{}_{\mu\nu}$ has a vanishing trace 
\begin{equation}\label{traceK}
K^\rho{}_{\mu\rho}=0\,.
\end{equation}

\subsection{Spin connection with intrinsic torsion}
Given the splitting of the torsion two-forms \eqref{T+T}, we split the spin connections accordingly, into parts carrying intrinsic and extrinsic torsion \footnote{Do not confuse the contorsion form $\kappa$ with the thermodynamic momentum introduced in Sec. \ref{sec:Thermal_Equilibrium}}
\begin{equation}\label{omegasdec}
\Omega= \mathring \Omega + \mathcal K\,,\qquad
\omega = \mathring \omega + \kappa \,,  
\end{equation}  
which, in an analogy with \eqref{OmegaABdef},  can be written in terms of tensors $\mathring \Omega^{AB}$ and $\mathcal K^{AB}$ as
\begin{equation}
\mathring \Omega^{AB}= \mathring \Omega \,\epsilon^{AB}+ \mathring \omega\, \varepsilon^{AB}\,,
\qquad
\mathcal K^{AB}= \mathcal K \,\epsilon^{AB}+ \kappa\,\varepsilon^{AB} \,.
\end{equation}
The extrinsic-torsion-free spin-connection satisfies
\begin{equation}\label{torsionfree-eqs}  
de^A + \mathring\Omega^A{}_B \wedge e^B = \mathring{ T}^A\,,
\end{equation}
whereas the extrinsic torsion two-form $\hat T^A$ can be expressed as
\begin{equation}
\hat T^A =  \mathcal K^A{}_B \wedge  e^B\,.
\end{equation}
Despite carrying intrinsic torsion, $\mathring\Omega$ and $\mathring \omega$ can be fully solved in terms of the tetrad and its derivatives. Indeed, using Eqs.~\eqref{TSigma} and \eqref{TKs}, leads to the following equations for the spin connection components
\begin{equation}\label{eqsforomegacirc}
\epsilon^{ab}\, \mathring \Omega^i=-\Sigma^{i[ab]}\,,
\quad
\varepsilon^{ij}\, 
 \omega^a=-
\Sigma^{a[ij]}\,,
\quad
\epsilon^a{}_{[b}  \,\mathring \Omega_{c]}=-\frac12 
\Sigma_{bc}{}{}^a\,,
\quad
\varepsilon^i{}_{[j}\, \mathring\omega_{k]}=-\frac12 
\Sigma_{jk}{}{}^i
\,,
\end{equation}
where we have introduced projections of the spin connection along the vierbein, i.e.,
\begin{equation}
\Omega_\mu = \Omega_A e^A{}_\mu,\qquad
 \Omega_A = E_A{}^\mu \Omega_\mu\,,
\end{equation}
and similarly for $\omega$. Solving Eq.~\eqref{eqsforomegacirc} leads to the following expression for the extrinsic-torsion-free spin connections 
\begin{subequations}\label{solcircomegas}
\begin{align}
\mathring\Omega_\mu&=-\epsilon_{ab}\left(
E^{a\alpha}\partial_{[\mu}e^b{}_{\alpha]}
-\frac12 E^{a\alpha}E^{b\beta}e_{c\mu}\partial_{\alpha}e^c{}_{\beta}
\right)\,,
\\
\mathring\omega_\mu&=\epsilon_{ij}\left(
E^{i\alpha}\partial_{[\mu}e^j{}_{\alpha]}
-\frac12 E^{i\alpha}E^{j\beta}e_{k\mu}\partial_{\alpha}e^k{}_{\beta}
\right)\,.
\end{align}
\end{subequations}

\subsection{Affine connection} 
It is possible to introduce an affine connection $\gamma^\rho{}_{\mu\nu}$ by defining a covariant derivative $\nabla_\mu$ through a vierbein postulate of the form
\begin{equation}\label{vpost}
\nabla_\mu e^A{}_\nu = \partial_\mu e^A{}_\nu + \Omega^{A}{}_{B\mu} \,e^B{}_\nu - \gamma^\rho{}_{\mu\nu} e^A{}_\rho=0\,.
\end{equation}
Such equations completely determine $\gamma^\rho{}_{\mu\nu}$, which can be written as
\begin{equation}
D_\mu \chi^A =e^A{}_\nu\nabla_\mu \chi^\nu\,,
\qquad \nabla_\mu \chi^\nu= E_A{}^\nu D_\mu \chi^A\;.
\end{equation}
Following \eqref{omegasdec}, we split the affine connection as
\begin{equation}
\gamma^\rho{}_{\mu\nu}=\mathring\gamma^\rho{}_{\mu\nu} + \kappa_{\mu\nu}{}{}^\rho\,, \label{gamma}
\end{equation}
where $\mathring \gamma^\rho{}_{\mu\nu}$ is the affine connection associated to $\mathring \Omega$ and $\mathring \omega$, whereas $\kappa_{\mu\nu}{}{}^\rho$ is the contorsion tensor associated to the contorsion forms $\mathcal K$ and $\kappa$. This means that the connection $\mathring\gamma^\rho{}_{\mu\nu}$ carries intrinsic torsion and $\kappa_{\mu\nu}{}^\rho$ is the contorsion tensor associated only with the extrinsic torsion, i.e., the part of the torsion that can be set to zero without imposing constraints on the vierbein. In order to find the explicit form of $\mathring\gamma^\rho{}_{\mu\nu}$ we replace Eqs.~\eqref{omegasdec} and \eqref{gamma} in \eqref{vpost} to find
\begin{equation}\label{defgammacircandkappa}
\mathring\gamma^\rho{}_{\mu\nu} = E_A{}^\rho \left(  \partial_\mu e^A{}_\nu + \mathring\Omega^A{}_{B\mu}  e^B{}_\nu \right)\,,
\qquad
 \kappa_{\mu\nu}{}^\rho 
 = \mathcal K^A{}_{B\mu} E_A{}^\rho e^B{}_\nu\,.
\end{equation}
After replacing the solution for the spin connection \eqref{solcircomegas} in \eqref{defgammacircandkappa}, the affine connection with vanishing extrinsic torsion can be written as
\begin{equation}\label{gammaintrinsic}
\begin{aligned}
\mathring \gamma^\rho{}_{\mu\nu}
&= \frac12 g^{\rho\lambda}\left( \partial_\mu g_{\lambda\nu}+ \partial_\nu g_{\mu\lambda}-\partial_\lambda g_{\mu\nu}\right)
\\&
+\frac12 h^{\rho\lambda}\left( \partial_\mu h_{\lambda\nu}+ \partial_\nu h_{\mu\lambda}-\partial_\lambda h_{\mu\nu}\right)
+ \left(g^{\rho\alpha} g_{\nu\lambda}+ h^{\rho\alpha} h_{\nu\lambda} \right) K^\lambda{}_{\mu\alpha}\,,
\end{aligned}
\end{equation}
where we have used the definition \eqref{Krhomunu}. The fact that $K^\rho{}_{\mu\nu}$ satisfies the traceless condition \eqref{traceK} can be used to show that the covariant divergence of a vector $V^\mu$ with respect to the extrinsic-torsion-free affine connection $\mathring\gamma^\rho{}_{\mu\nu}$ satisfies the usual property
\begin{equation}\label{covdiv}
(   \mathring\nabla_\mu -\mathring T^\rho{}_{\rho\mu})V^\mu= \partial_\mu V^\mu +\mathring\gamma^\rho{}_{\mu\rho} \,V^\mu
= \frac1{|e|} \partial_\mu \left(|e|\; V^\mu\right) \,.
\end{equation}
which guarantees that the divergence theorem holds in this type of geometry even though $\mathring \gamma^\rho{}_{\mu\nu}$ is not the standard Levi-Civita connection.

\section{Effective action and conserved currents}

Previous construction facilitates coupling the multi-Weyl fermions to external gauge fields and geometry, the minimally coupled theory has action
\begin{equation}\label{eq_Curved_Action}
    S = \int \, \Re\,\bar\phi\left[i \slashed D_\parallel + \,m^{1-n} (i\slashed D_\perp)^n \right]\phi\,dV \,,
\end{equation}
where
\begin{equation}
\slashed D_\parallel \phi= E_a{}^\mu \gamma^a D_\mu \phi, \qquad
\slashed D_\perp= E_i{}^\mu \gamma^i D_\mu \phi, \qquad \gamma^A= (\sigma^1,-i\sigma^2,\mathbb \sigma^0,-i\sigma^3)\,,
\end{equation}
 and $D_\mu \phi$ is given in Eq.~\eqref{Dvectorfermion}. The energy-momentum, spin, angular, and U(1) current operators are defined from the variation of the action with respect to the external fields
\begin{equation}
\d S=\int_{\mathcal{M}}  \left(\d e^A\wedge\star\mathcal  T_{ A} +  \d\Omega\wedge \star\mathcal S + \d\omega\wedge \star\mathcal J + \d A\wedge \star J   \right)\,,
\end{equation}
in particular the current operators for the system read
\begin{subequations}
\begin{align}
\star \mathcal T\,_A &= -Re\left[i\bar\phi\,\star\gamma_\parallel D_A\phi +  m^{1-n}i^n\sum_{j=1}^n (-1)^{j-1}  \bar\phi (\overleftarrow{\slashed D_\perp})^{j-1}\star\gamma_\perp\,D_A\slashed D_\perp^{n-j}\phi\right] \nonumber \\
& \qquad\qquad\qquad\qquad\qquad\qquad\qquad\qquad\qquad\qquad\qquad\quad+ \mathcal L (dS)_A\,,\\
\star \mathcal S &= \frac{1}{2}Re\left[i\bar\phi\,\star\gamma_\parallel\sigma^3 \phi  + m^{1-n}i^{n}\sum_{j=1}^n (-1)^{j-1}\bar\phi (\overleftarrow{\slashed D_\perp})^{j-1}\star\gamma_\perp\sigma^3\slashed D_\perp^{n-j}\phi\right]\,,\\
\star \mathcal J &= -\frac{n}{2}Re\left[\bar\phi\,\star\gamma_\parallel\sigma^3 \phi - m^{1-n}i^{n+1}\sum_{j=1}^n (-1)^{j-1}\bar\phi (\overleftarrow{\slashed D_\perp})^{j-1}\star\gamma_\perp\sigma^3\slashed D_\perp^{n-j}\phi\right]\,,\\
\star J &= -Re\left[\bar\phi\,\star\gamma_\parallel \phi - m^{1-n}i^{n+1}\sum_{j=1}^n (-1)^{j-1}  \bar\phi (\overleftarrow{\slashed D_\perp})^{j-1}\star\gamma_\perp\slashed D_\perp^{n-j}\phi\right]\,,
\end{align}
\end{subequations}
where we have defined $\star\gamma = \gamma^A(dS)_A$.  On the other hand, their expectation values can be obtained from the generating functional $\mathcal Z[e^A,\Omega^{AB},A]$ defined as
    \begin{equation}
\mathcal Z[e^A,\Omega^{AB},A] = -\log\int D\bar\Psi D\Psi e^{-S_E[\bar\Psi,\Psi;e^A,\Omega^{AB},A]}\,,
\end{equation}  
which under gauge variation with parameters $\chi$ will satisfy
\begin{equation}\label{eq_variation_action}
\delta_\chi\mathcal Z = W_{\text{anom}}[e^A,\Omega^{AB},A;\chi]\,,
\end{equation}
with $W_{\text{anom}}$ containing the gauge and gravitational anomalies of the system, nonetheless a detailed analysis of the hydrodynamic description including anomalous contributions escapes the scope of this paper and will be left for future studies, therefore, the existence of $W_{\text{anom}}$ will be ignored.

The Ward identities of the system can be obtained by explicitly evaluating the left-hand-side of Eq.~\eqref{eq_variation_action} which takes the form
\begin{equation}\label{eq:vareffact2}
\d_\chi \mathcal Z=-\int_{\mathcal{M}}  \left(\xi^A\frac{\d \mathcal Z}{\d \xi^A}  + \bar\lambda^{(S)} \frac{\d \mathcal Z}{\d  \bar\lambda^{(S)}} + \bar\lambda^{(L)} \frac{\d \mathcal Z}{\d  \bar\lambda^{(L)}} + \bar\lambda^{(Q)}\frac{\d \mathcal Z}{\d  \bar\lambda^{(Q)}}   \right)+\int_{\mathcal{ M}}d\star\tilde N\,,
\end{equation}
with $\tilde N = \mathcal T\,_A \xi^A + \mathcal S\bar \lambda^{(S)} +  \mathcal{J} \bar\lambda^{(L)} + J\bar\lambda^{(Q)}$. Therefore, each term in Eq.~\eqref{eq:vareffact2} must vanish independently and the Ward identities read
\begin{subequations}\label{eq_WI_scale}
\begin{align}
D\star\mathcal T\,_A -T_A^B\wedge\star\mathcal T\,_B -\mathcal{ F}_A^{(S)}\wedge\star\mathcal S -\mathcal{ F}_A^{(L)}\wedge\star\mathcal J  &= F_A\wedge\star J \,, \\
 d\star\mathcal S  &=  \star\mathcal T\,^A\wedge \epsilon_A \,, \\
 d\star\mathcal J &=  \star\mathcal T\,^A\wedge \varepsilon_A \,, \\
 d\star J &= 0 \,,
\end{align}
\end{subequations}
where we have used \eqref{defdSA}, the gauge field transformations \eqref {Tgf1}, and defined the one-form
\begin{equation}
\mathcal F_A \equiv \iota_{E_A} \mathcal F =  T^B_A P_B  
+ \frac{1}{2}\left( 
\mathcal F^{(S)}_A \,\epsilon^{BC}+ \mathcal F^{(L)}_A\, \varepsilon^{BC} 
\right) M_{BC} + F_A\,Q\,,
\end{equation}
in terms of the $SO(1,1)\times SO(2)$ curvature \eqref{CurvatureTwo-form}. Moreover, if the system possesses a symmetry generated by Killing parameters $\chi_K=(\xi_K,\lambda_K^{AB},\lambda_K^{(Q)})$ we can interpret the effective action as the hydrostatic partition function of the system
\begin{equation}\label{eq:partfunct}
\mathcal Z = \int_\mathcal M \star \mathcal P\,,
\end{equation}
where $\mathcal P$ is a local function of the spacetime coordinates that we interpret as the hydrostatic pressure. Therefore, after equating \eqref{eq:vareffact2} with the variation of \eqref{eq:partfunct} and using the equations of motion we obtain that the Noether current
\begin{equation}
N=\mathcal T\,_A \xi_K^A + \mathcal S\bar \lambda_K^{(S)} +  \mathcal{J} \bar\lambda_K^{(L)} + J\bar\lambda_K^{(Q)} - \mathcal P\xi_K\,, \label{Noether}
\end{equation}
must be conserved, i.e.
\begin{equation}\label{conservationofN}
d\star N=0\,.
\end{equation}

\section{Thermal equilibrium and partition function}\label{sect:mW_PartitionFunction}

In order to have the notion of hydrostatic equilibrium in the system, a set of Killing parameters are needed on the manifold such that the system can reach a steady regime with the external forces and local hydrodynamic variables compensating each other.  Given Killing parameters $\chi_K=(\xi_K,\lambda_K^{AB},\lambda_K^{(Q)})$ with $\xi_K$  future-oriented, in what follows we will  define the hydrostatic variables.

First we note that the projectors $g^A_B,h^A_B$ allow one to decompose  the tangent spaces as $T_p\mathcal M = \mathbb R^{1,1}_\parallel\oplus \mathbb R^2_\perp$. In particular, using the vector field $\xi_K$ we can construct vectors $u^A,v^A$ satisfying
 \begin{equation}
 u_A u^A = -1\,, \qquad v_A v^A = 1\,, \qquad u_A v^A = 0\,,
 \end{equation}
 belonging to the longitudinal and transverse subspaces in $T_p\mathcal M$ respectively, and defined as
\begin{equation}
\b u^A=g^A_{B}\xi_K^B\,,\quad \quad \bar\a \, v^A=\, h^A_{B}\xi_K^B\,.
\end{equation}
 with the inverse temperature $\beta=T^{-1}$ and (reduced) momentum's chemical potential $\bar\alpha=\beta\a$ defined as 
 \begin{equation}
 \b^{2} = - g(\xi_K,\xi_K)\,,\qquad \bar \a^2  = h(\xi_K,\xi_K)\,.
 \end{equation}
 The fluid velocity can be expanded as
 \begin{equation}
 U^A = u^A + \a\, v^A\,.
 \end{equation}
 At this point we find it convenient to introduce the projectors
 \begin{subequations}
 \begin{align}
     \Delta^A\,_B &=  \d^A_B + U^Au_B\,,\\
     P^A\,_B &= \Delta^A\,_B - v^A\hat v_B\,,
\end{align}
\end{subequations}
 where $\hat v_A = v_A+\a u_A$, such that
  $\Delta^A\,_B$ projects transverse to the fluid velocity, whereas $P^A\,_B$ projects transverse to both the fluid velocity and the momentum's chemical potential ($\a v^A$). 
 In addition, we define also the boost-rotations, and $U(1)$ (reduced) chemical potentials as
\begin{equation}
  \bar\mu^{AB} = \bar\mu^{(S)}\epsilon^{AB}+\bar\mu^{(L)}\varepsilon^{AB}\equiv \bar \lambda^{AB}_K\,,\qquad  \bar\mu \equiv 
  \bar\lambda^{(Q)}_K\,.
\end{equation}
Note that the structure of the tangent space and the properties of the degenerate epsilon tensors $\epsilon_{AB},\varepsilon_{AB}$ allow one to construct out of the Killing vector $\xi_K$, a complete set of vectors $(U,\tilde U,\tilde V,V)$ and dual co-vectors $(u,\tilde u,\tilde v,\hat v)$ defined as follows
\begin{subequations}
\begin{align}
&u = u^A e_A\,,\quad \tilde u = u^A\epsilon_A \,,\quad \tilde v = v^A\varepsilon_A\,,\quad   \hat v = v+\a u\,,\\
&\tilde U = g^{-1}\tilde u\,, \quad \tilde V = h^{-1}\tilde v\,, \quad V = h^{-1} v\,,\quad v = v^Ae_A.
\end{align}
\end{subequations}
with $g^{-1},h^{-1}$ the degenerate inverse `metrics'. On this basis, both $g$ and $h$ take the form
 \begin{equation}
g = -u\otimes u + \tilde u\otimes \tilde u\,,\qquad h =  v\otimes v + \tilde v\otimes \tilde v.
 \end{equation}

The hydrostatic relations are derived by expressing the Killing conditions $\delta_{\chi_K}\mathcal{A}=0$, with the components of $\delta_\chi \mathcal A$ given in Eq.~\eqref{Tgf1}, in terms of the hydrodynamic variables. In particular, the variation of the vierbeins $e^A$ implies that
\begin{subequations}
\begin{align}
\label{eq_Lie_g}
\mathcal L_{\xi_K} g  &= 0\,,\\
\label{eq_Lie_h}\mathcal L_{\xi_K} h   &= 0\,.
\end{align}
\end{subequations}
Since $g$ and $h$ are written in terms of the 1-forms $u,\tilde u,v, \tilde v$ we compute first their Lie derivatives
\begin{subequations}
\begin{align}
\mathcal L_{\xi_K} u &= \beta a^0 -d\beta \,,&\qquad a^0&=\iota_U du\,,\\
\mathcal L_{\xi_K} \tilde u &= \beta a^1 \,,&\qquad a^1&=\iota_U d\tilde u\,,\\
\mathcal L_{\xi_K} \tilde v &= \beta a^2 \,,&\qquad a^2&=\iota_U d\tilde v\,,\\
\mathcal L_{\xi_K} v &= \beta a^3 +d\bar\a \,, &\qquad a^3&=\iota_U dv\,.
\end{align}
\end{subequations}
Using these we can find the Lie derivative of the `metrics' as follows
\begin{subequations}
\begin{align}
   & \mathcal{L}_{\xi_K} g= \left[-\beta \left(u \otimes a^0+a^0 \otimes u\right)+u\otimes d\beta+d\b\otimes u+\beta \left( \tilde u \otimes a^1+a^1\otimes \tilde u \right) \right], \\
   & \mathcal{L}_{\xi_K} h= \left[ \b \left( v \otimes a^3+a^3 \otimes v\right)+v \otimes d\bar \a+d\bar\alpha\otimes v+\b \left( \tilde v \otimes a^2+a^2 \otimes \tilde v\right) \right].
\end{align}
\end{subequations}
Therefore, the hydrostatic conditions are obtained from all the independent projections of Eqs. \eqref{eq_Lie_g} and \eqref{eq_Lie_h} respectively. In particular, the scalar projections $\mathcal L_{\xi_K} g\,(U,U)$, $\mathcal L_{\xi_K} h\,(U,U)$, $\mathcal L_{\xi_K} g\,(U,V)$, $\mathcal L_{\xi_K} h\,(U,V)$, $\mathcal L_{\xi_K} g\,(V,V)$ and  $\mathcal L_{\xi_K} h\,(V,V)$ imply
\begin{subequations}
\begin{align}
    \iota_U d\beta &= 0\,, & \iota_U d\a &= 0\,,\\
      \iota_V(\beta a^0 -d\beta) &=0 \,,  & \iota_V(\beta a^3 +d\bar\a ) &=0\,.
\end{align}
\end{subequations}
On the other hand, the vector projectors are obtained from $\mathcal L_{\xi_K} g\,(U,\eta)$, $\mathcal L_{\xi_K} h\,(U,\eta)$, $\mathcal L_{\xi_K} g\,(V,\eta)$, $\mathcal L_{\xi_K} h\,(V,\eta)$ with $\eta$ a vector transverse to both $U,V$, i.e. $P^\mu_{\ \nu}\eta^\nu=\eta^\mu$
\begin{subequations}
\begin{align}
(\beta a^0 -d\beta) (\eta)&=0 \,, & (\beta a^3 +d\bar\a ) (\eta)&=0 \,,\\
\iota_V a^1 &= 0\,, &  \iota_Va^2 &= 0\,.
\end{align}
\end{subequations}
The tensor equations finally read
\begin{subequations}
\begin{align}
(a^1\otimes \tilde u +\tilde u\otimes a^1) (\eta,\gamma)&=0 \,,\\
 (a^2\otimes\tilde v +\tilde v\otimes a^2 ) (\eta,\gamma) &=0 \,,
\end{align}
\end{subequations}
with $\eta$ and $\gamma$ transverse vectors. The invariance of the gauge fields under the Killing transformation on the other hand implies
\begin{subequations}
\begin{align}
 \iota_U F + T d\bar\m  &= 0 \,,\\
 \iota_U \mathcal F^{AB} + T d\bar\mu^{AB}  &= 0\,.
 \end{align}
\end{subequations}
Therefore, in components, the hydrostatic relations read
\begin{subequations}
\begin{align}
U^\m\partial_\m T = U^\m\partial_\m \a &= 0\,,\\
U^\m\partial_\m \bar\mu = U^\m\partial_\m \bar\m^{AB} &= 0\,,\\
 E_\mu + T \Delta^\n\,_\m \partial_\n\bar\m  &= 0 \,,\\
 \mathcal E_\mu\,^{AB} + T \Delta^\n\,_\m\partial_\n\bar\mu^{AB}  &= 0\,,\\
T (a^0)_\m  + \Delta^\n\,_\m\partial_\n T &=0 \,, \\
  (a^3)_\m + T\Delta^\n\,_\m\partial_\n \bar\a & =0 \,,\\
  (a^1)_{\m} =  (a^2)_{\m} &=0 \,,
\end{align}
\end{subequations}
with the acceleration and electric fields defined as 
\begin{subequations}
\begin{align}
    (a^0)_\m &= U^\n\mathring\nabla_\n u_\m + u_\a\mathring {\mathcal G}^\a\,_{\m}\,,\\
      (a^1)_\m &= U^\n\mathring\nabla_\n \tilde u_\m + \tilde u_\a\mathring {\mathcal G}^\a\,_{\m}\,,\\
    (a^2)_\m &= U^\n\mathring\nabla_\n \tilde v_\m + \tilde v_\a\mathring {\mathcal G}^\a\,_{\m}\,,\\
    ( a^3)_\m &= U^\n\mathring\nabla_\n  v_\m +  v_\a\mathring{\mathcal G}^\a\,_{\m}\,,\\
    E_\m &=U^\n F_{\n\m}\,,\\
    \mathcal E_\m\,^{AB} &=U^\n \mathcal F_{\n\m}\,^{AB}\,,
\end{align}
\end{subequations}
and $\mathcal G^\alpha{}_\mu= T\bar{\mathcal G}^\alpha{}_\mu$. On the other hand, the rotations and boost chemical potentials must obey the relations
\begin{subequations}
\begin{align}
\mu^{(L)} - \frac{1}{2}T\mathring \nabla_\mu(\bar\a \tilde v^\mu) &=  U^\m \kappa_\mu\,,\\
\mu^{(S)} + \frac{1}{2}T\mathring \nabla_\mu(\beta\tilde u^\mu) &= U^\m \mathcal K_\mu \,.
\end{align}
\end{subequations}
With the notion of equilibrium being established, now we discuss the zeroth-order hydrodynamics. As we discussed before, the zeroth-order hydrostatic partition function must be a functional of the scalars constructed with no derivative. There are five zeroth-order scalars that characterize the hydrostatic system: the Lorentzian norm of the Killing vector $(\xi_K^A)$ defines a temperature $T$, whereas the Euclidean norm defines a chemical potential $\a$. On top of these, the system possesses three  chemical potentials: $\mu^{(S)}$ and  $\mu^{(L)} $ corresponding to boost-rotation charges, and $\mu$ corresponding to the U$(1)$ charge. 
Combining all these together, at the zeroth order the hydrostatic partition function is given by: 
\begin{equation}
\mathcal{Z} = \int  \mathcal P\left(T,\alpha,\mu^{(S)}, \mu^{(L)}, \mu\right)dV\,.
\end{equation}
As we have argued before, this partition function acts as the generating functional from which we can obtain various currents and the stress-energy tensor.  After varying with respect to different background fields, the ideal constitutive relations are
\begin{subequations}
\begin{align}
& J^\m = \rho U^\m \,\\
& \mathcal{S}^\m  =\rho^{(S)} U^\m ,\\
& \mathcal{ J}^\m  =  \rho^{(L)} U^\m ,\\
& \mathcal T^\mu\,_A = (\tilde{\mathcal{E}} u_A + p \hat v_A)U^\mu+ \mathcal P\Delta^\m\,_A. 
\end{align}
\end{subequations}
where we have defined
\begin{align}
s &= \frac{\partial\mathcal P}{\partial T}\, \quad p=\frac{\partial\mathcal P}{\partial \alpha},\quad    \rho^{(S)} = \frac{\partial\mathcal P}{\partial\mu^{(S)}}\,,\quad   \rho^{(L)} = \frac{\partial\mathcal P}{\partial\mu^{(L)}}\,,\quad    \rho= \frac{\partial\mathcal P}{\partial\mu}, 
\end{align}
 and the Gibbs-Duhem relation Eq.~\eqref{eq_Gibbs-Duhem2} 
 that incorporates the spin densities $\tilde{\mathcal{E}}+\mathcal P=Ts  +\rho \mu +\rho^{(S)} \mu^{(S)} +\rho^{(L)} \mu^{(L)}$ has been used. Moreover, the entropy current is  obtained by inserting the constitutive relations into  Eq.~\eqref{Noether}, which after straightforward simplifications takes the form
 \begin{equation}
 N^\mu =-s U^\mu.
 \end{equation}

\section{Discussion and Outlook}\label{sect:discussion}

In this work, we have developed a covariant framework to describe the hydrostatic equilibrium of mWSMs, which are characterized by their anisotropic dispersion relations and  topological charge. Our formalism, rooted in the gauging of the reduced spacetime symmetry group \((SO(1,1) \times SO(2)) \ltimes \mathbb{R}^4\), leads to a non-Lorentzian geometry tailored to capture the physical behavior of these systems. A pivotal achievement of this study is the construction of a new class of non-Lorentzian curved spacetime background that naturally incorporates the anisotropic properties of mWSMs, facilitated through the coupling of multi-Weyl fermions to external gauge and geometric fields.
One of the notable features emerging from our analysis is that these spacetimes are intrinsically torsionful, forcing a distinction between intrinsic and extrinsic torsion components in the underlying geometry. 

Our derivation of the hydrostatic partition function at zeroth order lays the groundwork for understanding non-dissipative transport in mWSMs. It demonstrates that despite the absence of Lorentz invariance, the system retains a robust structure enabling the definition of equilibrium thermodynamic variables such as temperature, chemical potentials, and pressure. These variables, through a carefully defined set of Killing parameters, ensure the compatibility of our geometric formalism with hydrostatic equilibrium conditions.
Furthermore, we emphasize that while our motivation stemmed from condensed matter realizations of mWSMs, the formalism possesses broader applicability. Specifically, it extends naturally to ultra-relativistic hydrodynamic systems such as Bjorken flow, which share a similar symmetry structure. This underscores the utility of our framework as a general tool for analyzing systems with reduced Lorentz symmetry in both high-energy and condensed matter physics.

Our work paves the road to the construction of a fully non-dissipative hydrodynamic theory for these systems. Moreover, it allows for a systematic analysis of anomaly-induced transport, especially the one related to expected gravitational anomalies of the system \cite{Dantas2020Non-AbelianSemimetals}. In materials such as Graphene and Weyl semimetals, straining manifests in the form of pseudo-gauge fields and generates an effective curved geometry \cite{Vozmediano2010GaugeGraphene, Cortijo2015ElasticSemimetals,Palumbo:2024vqv}; it would be interesting to understand how these phenomena generalize to the case of mWSMs. In fact, recent studies find that pseudo-gauge fields do not emerge from strain in mWSMs \cite{SubramanyanGeometricSemimetals, SukhachovElectronicSystems}, therefore, we may wonder whether all the effect of strain fields is encapsulated in the geometric fields of our theory. 

Furthermore, our formalism is applicable to exploring out-of-equilibrium dynamics via the Schwinger-Keldish effective action formalism \cite{Liu2017LecturesHydrodynamics, Akyuz:2023lsm} or using gauge/gravity duality \cite{Glorioso:2018mmw, Ghosh:2020lel, Ghosh:2022fyo, Baggioli:2023tlc, Baggioli:2024zfq, Bu:2025zad}, which presents an exciting avenue. Ultimately, we anticipate that the symmetry-based approach adopted here will catalyze further developments in the hydrodynamic modeling of anisotropic and non-boost-invariant systems.

%%%%%%%%%%%%%%%%%%%%%%%%%%%%%%%%%
%%%%%%%%%%%%%%%%%%%%%%%%%%%%%%%%%
%%%%%%%%%%%%%%%%%%%%%%%%%%%%%%%%%
%%%%%%%%%%%%%%%%%%%%%%%%%%%%%%%%%
\acknowledgments
We thank E. Bergshoeff, Kristan Jensen and M. Arshad Momen for enlightening comments and discussions.  F. P.-B. and P.S.-R have been funded by the Norwegian Financial Mechanism 2014-2021 via the Narodowe Centrum Nauki (NCN) POLS grant 2020/37/K/ST3/03390. P.S.-R. has been supported by a Young Scientist Training Program (YST) fellowship at the Asia Pacific Center for Theoretical Physics (APCTP) through the Science and Technology Promotion Fund and the Lottery Fund of the Korean Government. P.S.-R. has also been supported by the Korean local governments in Gyeongsangbuk-do Province and Pohang City.

%%%%%%%%%%%%%%%%%%%%%%%%%%%%%%%%%
%%%%%%%%%%%%%%%%%%%%%%%%%%%%%%%%%
%%%%%%%%%%%%%%%%%%%%%%%%%%%%%%%%%
%%%%%%%%%%%%%%%%%%%%%%%%%%%%%%%%%
%%%%%%%%%%%%%%%%%%%%%%%%%%%%%%%%%
%%%%%%%%%%%%%%%%%%%%%%%%%%%%%%%%%
%%%%%%%%%%%%%%%%%%%%%%%%%%%%%%%%%
%%%%%%%%%%%%%%%%%%%%%%%%%%%%%%%%%
%%%%%%%%%%%%%%%%%%%%%%%%%%%%%%%%%

\bibliographystyle{JHEP}
\bibliography{refmanual}
\end{document}